# Who Leads in the Shadows? ERGM and Centrality Analysis of Congressional Democrats on Bluesky

Gordon Hew[1] and Ian McCulloh[1][0000-0003-2916-3914]

[1] Johns Hopkins University, Baltimore MD 21218, USA
imccull4@jhu.edu

**Abstract.** Following the 2024 U.S. presidential election, Democratic lawmakers and their supporters increasingly migrated from mainstream social media platforms like X (formerly Twitter) to decentralized alternatives such as Bluesky. This study investigates how Congressional Democrats use Bluesky to form networks of influence and disseminate political messaging in a platform environment that lacks algorithmic amplification. We employ a mixed-methods approach that combines social network analysis, exponential random graph modeling (ERGM), and transformer-based topic modeling (BERTopic) to analyze follows, mentions, reposts, and discourse patterns among 182 verified Democratic members of Congress. Our findings show that while party leaders such as Hakeem Jeffries and Elizabeth Warren dominate visibility metrics, overlooked figures like Marcy Kaptur, Donald Beyer, and Dwight Evans occupy structurally central positions, suggesting latent influence within the digital party ecosystem. ERGM results reveal significant homophily along ideological, state, and leadership lines, with Senate leadership exhibiting lower connectivity. Topic analysis identifies both shared themes (e.g., reproductive rights, foreign conflicts) and subgroup-specific issues, with The Squad showing the most distinct discourse profile. These results demonstrate the potential of decentralized platforms to reshape intra-party communication dynamics and highlight the need for continued computational research on elite political behavior in emerging digital environments.

## 1    Introduction

In the wake of heightened political polarization in the United States, the migration of users across social media platforms has become a critical frontier for both public discourse and computational social science. Following the 2024 U.S. presidential election, significant shifts in platform allegiance emerged—particularly among Democratic political figures and their supporters—prompting renewed interest in decentralized and algorithmically transparent alternatives to mainstream platforms such as X (formerly

Twitter) (Ambrose, 2024; NPR, 2024). Among these, Bluesky, a federated microblogging service built on the open-source AT Protocol, has experienced rapid growth, reaching over 35 million users by April 2025 (Bluesky, n.d.).

Unlike X, Bluesky emphasizes decentralization, user autonomy, and content curation through customizable feeds and moderation services (Kleppmann et al, 2024). Its open API and lack of algorithmic amplification make it an appealing testbed for researchers interested in organic network formation, elite communication strategies, and the self-organization of political communities online. Despite its growing political relevance, relatively little academic research has systematically explored Bluesky's network topology—particularly among verified or high-profile users such as U.S. Congressional Democrats.

This paper leverages exponential random graph modeling (ERGM) to analyze the structural determinants of connectivity among Democratic members of Congress active on Bluesky. We specifically examine how network ties—such as follows, mentions, and reposts—are influenced by political group membership, leadership status, and ideological alignment. Additionally, we analyze topic diffusion and clustering across factions within the Democratic Party, investigating whether content differs significantly between moderates, progressives, and party leadership.

We pose the following research questions:

- **RQ1**: What endogenous and exogenous structural features (e.g., mutual follows, ideological similarity, leadership roles) explain the formation of ties among Congressional Democrats on Bluesky?
- **RQ2**: To what extent do topical interests differ between factions within the Democratic network, and how do these differences manifest in observable communication patterns?
- **RQ3**: To what extent does structural centrality (e.g., degree, betweenness, eigenvector centrality) within the Bluesky network of Congressional Democrats correlate with public perceptions of influence as reflected in national popularity polls, media coverage, or leadership roles within the Democratic Party?

By addressing these questions, we aim to contribute to a growing body of work on political communication networks, particularly within emergent or decentralized platforms. This study not only highlights the early adoption patterns and discourse trends of influential political actors but also provides a replicable methodological framework for social network modeling in evolving digital spaces.

## 2    Background

The 2024 U.S. Presidential election marked a significant shift in the digital landscape of political discourse, particularly among Democratic political figures and their supporters. In the aftermath of the election, there was a notable migration from traditional platforms like X (formerly Twitter) to alternative, decentralized platforms such as Bluesky. Despite its growing prominence, scholarly research on Bluesky remains limited. Most existing studies focus on its technical architecture and comparisons with established platforms. For instance, Buzelin et al. (2025) analyzed political discourse on Discord during the 2024 U.S. Presidential election, highlighting the migration of political discussions to alternative platforms. However, their study did not delve into network structures or employ social network analysis methods, leaving a gap in understanding the structural aspects of political interactions on emerging platforms like Bluesky.

In contrast, extensive research has been conducted on political interactions on Twitter. Chamberlain et al. (2021) analyzed over a decade of Twitter interactions among U.S. Congress members, employing community detection algorithms to infer political affiliations and measure ideological leanings. Their study highlighted the potential of social media data in understanding political networks. Similarly, Sadayappan et al. (2018) utilized Exponential Random Graph Models (ERGMs) to evaluate political party cohesion, revealing that factors like gender and ideological similarity significantly influence cross-party interactions. This approach underscores the value of ERGMs in dissecting the nuances of political networks.

Social Network Analysis (SNA) offers a robust framework for understanding the complexities of political interactions on social media platforms (McCulloh et al, 2013). Centrality measures—such as degree, betweenness, closeness, and eigenvector centrality—provide insights into the influence and importance of individuals within a network . Notably, individuals with high centrality may not always be the most visible or popular figures; instead, their strategic positions within the network enable them to wield significant influence, often remaining unnoticed by mainstream media.

Understanding community structures within political networks is equally crucial. Community detection methods, help identify clusters or factions within a network, shedding light on the underlying ideological divisions and alliances (Ozer et al., 2016). In the context of the Democratic Party, analyzing these community structures post-2024 election can reveal insights into emerging factions, core issues, and the evolving dynamics that define the party's internal landscape.

This study aims to bridge the existing research gap by focusing exclusively on Bluesky. By employing ERGMs and SNA methods, we seek to analyze the network structures among Democratic members of Congress on the platform, identify influential actors based on centrality measures, and uncover prevailing themes through topic modeling. This approach will provide a comprehensive understanding of the factors contributing to connectivity and influence within the Democratic Party's digital ecosystem on Bluesky.

## 3    Methods

### 3.1    Data Collection

A complete list of current members of the U.S. House of Representatives and Senate was collected from the official House directory (U.S. House of Representatives, 2024) and Senate website (U.S. Senate, 2024) using the BeautifulSoup Python library. The scraped data included names, chamber affiliation, and states, which were stored in a structured CSV file for reproducibility and used to assign node-level attributes in subsequent network modeling.

To map each congressional Democrat to their Bluesky profile, we first extracted verified account handles listed on the House Democrats website and the Senate Democrats "Starter Pack" Bluesky post. For Senators not listed in the Starter Pack, public accounts were manually verified using name, image, and domain association (e.g., house.gov, senate.gov). In cases of multiple accounts, preference was given to those officially listed by the caucus, ensuring consistent mapping to formal political identities.

Congressional Democrats were manually assigned to formal and informal groups using public membership directories. These include Democratic House and Senate leadership, the Congressional Progressive Caucus, Blue Dog Coalition, New Democrat Coalition, the Problem Solvers Caucus, and The Squad. Each member may belong to multiple groups. This many-to-many relationship was stored in a separate CSV and joined to the master file via unique identifiers.

We used the `atproto` Python library to collect each member's network and post data from Bluesky. For each verified account: The follows list was retrieved and converted into a binary directed adjacency matrix; Mentions and reposts were extracted from post histories and converted into weighted adjacency matrices based on frequency. Additionally, full post histories were collected (text, date, mentions, and repost metadata), stored in JSON and CSV formats, and cleaned for downstream semantic analysis.

Out of 211 House Democrats and 47 Senate Democrats (including Independents who caucus with the party), 182 had identifiable Bluesky accounts. Missing nodes are acknowledged as a potential bias; we perform sensitivity analysis by simulating random removals of 5–10% of nodes to evaluate effects on network metrics

### 3.2 Network Analysis

We construct three directed graphs: one each for follows, mentions, and reposts. Each graph uses Congressional Democrats as nodes, and interactions as edges. Weighted edges reflect frequency (e.g., number of mentions). To understand structural influence, we compute: Degree centrality (in-degree and out-degree); Closeness and betweenness centrality, capturing brokerage roles and efficiency of access (Freeman, 1979); Eigenvector centrality, identifying well-connected and prestigious actors (Bonacich, 1987). These measures are particularly important for identifying influential but non-prominent figures who may lack public recognition but are structurally vital—a key distinction highlighted in social media network studies (McCulloh, Armstrong, & Johnson, 2013; Himelboim et al., 2013).

To model tie formation, we construct separate ERGMs for follows, mentions, and reposts using the statnet suite in R. ERGMs allow modeling of both endogenous structural effects (e.g., reciprocity, transitivity) and exogenous nodal covariates such as: Chamber (Senate or House); Group membership (e.g., Progressive Caucus); Leadership role; Gender; Ideological affiliation (as a categorical variable from group membership). Model selection is guided by AIC and convergence diagnostics. This approach builds on prior work that identified gender and ideology as significant predictors of cross-party ties in legislative networks (Sadayappan, McCulloh, & Piorkowski, 2018).

### 3.3 Topic Modeling

To analyze discourse trends, we apply BERTopic, a transformer-based topic modeling framework that extracts coherent, semantically-rich topics using class-based TF-IDF embeddings (Grootendorst, 2022). Posts are preprocessed using lowercasing, stopword removal, lemmatization, and emoji stripping.

Topics are modeled separately for each subgroup and the overall corpus. The top eight topics per group are vectorized using TF-IDF and compared using cosine similarity to assess alignment or divergence in discourse across ideological lines. Topic coherence is evaluated using the UMass score, and low-quality or redundant topics are filtered.

## 4 Findings

### 4.1 Analysis of Followers Network

The Congressional Democrats Follows Network in Figure 1 has 219 nodes and 15,015 directed edges, with a moderate density of 0.3145 and diameter of 5. This level of connectivity reflects a densely intertwined set of online ties within the Democratic Party's Bluesky presence. As anticipated, the network clusters strongly by chamber membership (House vs. Senate), a pattern statistically validated by Exponential Random Graph Modeling (ERGM) results.

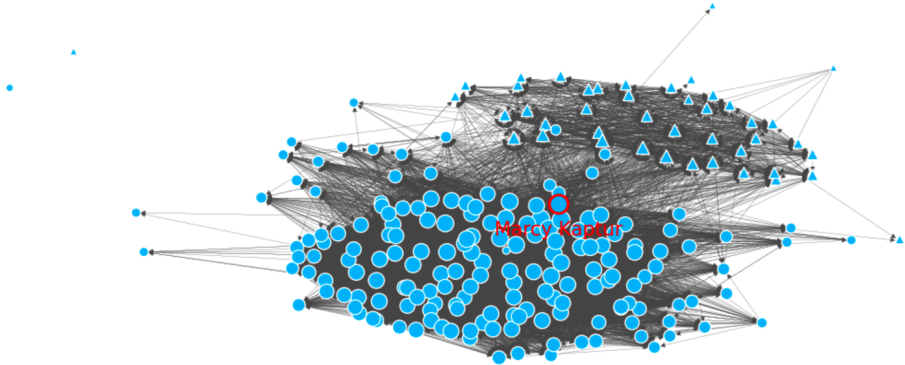

**Fig. 1.** Congressional Democrats Bluesky Follows Network Visualization.

One of the most intriguing findings is the centrality of Representative Marcy Kaptur (OH), who possesses the highest closeness centrality (0.9495) and highest betweenness centrality (0.0786) despite lacking formal affiliation with any major caucus or leadership group. Kaptur's structural role suggests she occupies a highly accessible and bridging position, making her a potential influence broker in the digital space, even though she is often overlooked in traditional political media coverage.

By contrast, Representative Katherine Clark (MA) and Representative Gwen Moore (WI) also rank highly on betweenness and closeness centrality but are already acknowledged leaders, affiliated with the Democratic House Leadership and Congressional Progressive Caucus, respectively.

ERGM analysis (Table 1) shows strong homophily effects: chamber membership, state, verified status, and most group affiliations significantly increase the likelihood of a tie. However, membership in the Democratic Senate Leadership has a negative coefficient, indicating these members are less likely to form reciprocal or follower ties—perhaps reflective of more hierarchical or siloed Senate communication patterns. Notably, The Squad and Congressional Progressive Caucus members are highly interconnected, consistent with prior studies of ideological clustering on social media (Chamberlain et al., 2021).

**Table 1.** Congressional Democrats Follower Network ERGM Results.

| | Estimate | Std. Error | MCMC % | z value | $Pr(>|z|)$ |
|---|---|---|---|---|---|
| edges | -3.87525 | 0.11860 | 0 | -32.675 | $< 1e - 04$ |
| mutual | 0.35303 | 0.03156 | 0 | 11.185 | $< 1e - 04$ |
| nodematch.member_type | 2.71894 | 0.04420 | 0 | 61.508 | $< 1e - 04$ |
| nodematch.state | 0.70633 | 0.04885 | 0 | 14.460 | $< 1e - 04$ |
| nodematch.is_verified | 0.08255 | 0.02111 | 0 | 3.911 | $< 1e - 04$ |
| nodematch.blue_dog_coalition | 0.11454 | 0.05441 | 0 | 2.105 | 0.035274 |
| nodematch.problem_solvers_caucus | 0.21310 | 0.02888 | 0 | 7.379 | $< 1e - 04$ |
| nodematch.new_democrat_coalition | 0.07295 | 0.02101 | 0 | 3.472 | 0.000517 |
| nodematch.democratic_senate_leadership | -0.13164 | 0.05643 | 0 | -2.333 | 0.019668 |
| nodematch.democratic_house_leadership | 0.10309 | 0.04815 | 0 | 2.141 | 0.032287 |
| nodematch.the_squad | 0.38593 | 0.06421 | 0 | 6.010 | $< 1e - 04$ |
| nodematch.congressional_progressive_caucus | 0.03977 | 0.02140 | 0 | 1.858 | 0.063124 |

This network uncovers that legislative seniority or media profile may not correlate with online embeddedness. Marcy Kaptur emerges as an unheralded yet structurally influential figure, exemplifying how social network metrics can spotlight overlooked actors (McCulloh, Armstrong, & Johnson, 2013).

### 4.2  Analysis of Mentions Network

The Mentions Network in Figure 2 is significantly sparser, with only 1,434 directed edges, a density of 0.0300, and a diameter of 10. The network lacks strong chamber-based clustering, suggesting that mentions may follow thematic or issue-based patterns rather than institutional roles.

Notably, Senator Cory Booker (NJ), Representative Hakeem Jeffries (NY), and Senator Elizabeth Warren (MA) all rank within the top five for both closeness and betweenness centrality, reflecting their dual roles as party leaders and public communicators. However, less prominent figures like Representative Delia Ramirez (IL) and Emanuel Cleaver (MO) also emerge as highly central in the mentions network, revealing that policy-specific discourse (e.g., housing, equity) may elevate certain voices even if they are less prominent in the follower graph.

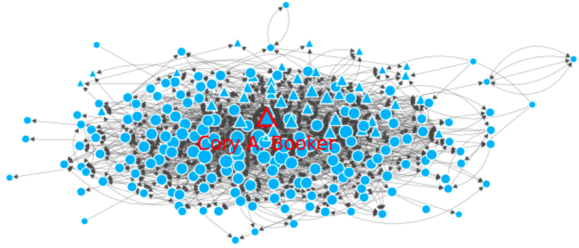

**Fig. 2.** Congressional Democrats Bluesky Mentions Network Visualization.

ERGM results (Table 2) reaffirm the importance of reciprocity and shared state affiliation in mentions. Interestingly, membership in Senate leadership negatively correlates with mentions, possibly due to these figures being less interactive and more broadcast-oriented. This contrast between formal leadership and interactive visibility highlights how mentions serve as a proxy for conversational or issue-based influence—often elevating progressive or policy-driven members outside traditional power hierarchies.

**Table 2.** Congressional Democrats Mentions Network ERGM Results.

|  | Estimate | Std. Error | MCMC % | z value | $Pr(>|z|)$ |
|---|---|---|---|---|---|
| sum | -3.25077 | 0.38417 | 0 | -8.462 | $< 1e-04$ |
| nonzero | -3.35241 | 0.09581 | 0 | -34.989 | $< 1e-04$ |
| mutual.min | 1.44118 | 0.11105 | 0 | 12.978 | $< 1e-04$ |
| nodematch.sum.member_type | 0.47024 | 0.06216 | 0 | 7.564 | $< 1e-04$ |
| nodematch.sum.state | 1.11855 | 0.05941 | 0 | 18.828 | $< 1e-04$ |
| nodematch.sum.is_verified | 0.22416 | 0.04744 | 0 | 4.725 | $< 1e-04$ |
| nodematch.sum.blue_dog_coalition | 1.19767 | 0.32032 | 0 | 3.739 | 0.000185 |
| nodematch.sum.problem_solvers_caucus | 0.31459 | 0.08439 | 0 | 3.728 | 0.000193 |
| nodematch.sum.new_democrat_coalition | 0.16556 | 0.04994 | 0 | 3.315 | 0.000915 |
| nodematch.sum.democratic_senate_leadership | -0.71027 | 0.06838 | 0 | -10.387 | $< 1e-04$ |
| nodematch.sum.democratic_house_leadership | -0.10635 | 0.09178 | 0 | -1.159 | 0.246571 |
| nodematch.sum.the_squad | -0.10315 | 0.12820 | 0 | -0.805 | 0.421070 |
| nodematch.sum.congressional_progressive_caucus | 0.19127 | 0.04937 | 0 | 3.874 | 0.000107 |

### 4.3 Analysis of Repost Network

The Reposts Network in Figure 3 is the sparsest of the three, with only 468 directed edges, a density of 0.0098, and a diameter of 13. This indicates that content amplification through reposting is far less common among Democratic Congress members than following or mentioning.

As in the mentions network, we observe a few dominant figures like Senator Elizabeth Warren, Senator Charles Schumer, and Representative Hakeem Jeffries, but also the emergence of Representative Dwight Evans (PA) and Donald Beyer (VA) as central nodes. Their content appears to resonate across ideological lines, prompting repeated amplification.

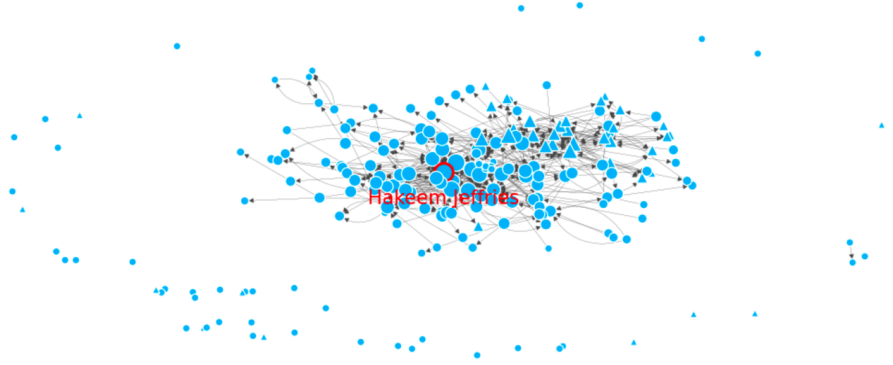

**Fig. 3.** Congressional Democrats Bluesky Repost Network Visualization.

The ERGM (Table 3) indicates that repost ties are more likely among members of the same state, ideological caucus, or chamber, with Democratic House and Senate leadership negatively associated with repost activity. This suggests that reposts are a more grassroots and peer-to-peer behavior, less likely among formal elites.

**Table 3.** Congressional Democrats Repost Network ERGM Results.

|  | Estimate | Std. Error | MCMC % | z value | $Pr(>|z|)$ |
|---|---|---|---|---|---|
| sum | -4.50874 | 0.77296 | 0 | -5.833 | $< 1e-04$ |
| nonzero | -3.97581 | 0.16203 | 0 | -24.538 | $< 1e-04$ |
| mutual.min | 1.50311 | 0.22931 | 0 | 6.555 | $< 1e-04$ |
| nodematch.sum.member_type | 1.36963 | 0.15733 | 0 | 8.706 | $< 1e-04$ |
| nodematch.sum.state | 0.87873 | 0.10279 | 0 | 8.548 | $< 1e-04$ |
| nodematch.sum.is_verified | 0.12677 | 0.07992 | 0 | 1.586 | 0.112676 |
| nodematch.sum.blue_dog_coalition | 1.62879 | 0.67626 | 0 | 2.409 | 0.016017 |
| nodematch.sum.problem_solvers_caucus | 0.69984 | 0.18134 | 0 | 3.859 | 0.000114 |
| nodematch.sum.new_democrat_coalition | 0.21394 | 0.08900 | 0 | 2.404 | 0.016228 |
| nodematch.sum.democratic_senate_leadership | -0.83193 | 0.12481 | 0 | -6.666 | $< 1e-04$ |
| nodematch.sum.democratic_house_leadership | -0.69802 | 0.11314 | 0 | -6.170 | $< 1e-04$ |
| nodematch.sum.the_squad | 0.03781 | 0.24550 | 0 | 0.154 | 0.877596 |
| nodematch.sum.congressional_progressive_caucus | 0.20618 | 0.08388 | 0 | 2.458 | 0.013975 |

This sparse network reinforces findings from Himelboim et al. (2013) that information diffusion in political networks is not solely driven by central elites, and that influential amplifiers often occupy intermediate or bridging roles. Beyer and Evans exemplify such figures: connected enough to receive reposts from multiple ideological factions, yet not always featured in mainstream political commentary

### 4.4 Summary of Network Findings

Together, these three networks paint a nuanced picture of intra-party digital behavior. While formal leaders dominate certain aspects of online visibility, overlooked members such as Marcy Kaptur, Dwight Evans, and Donald Beyer hold structurally significant positions that make them key intermediaries in the flow of influence and discourse.

By leveraging ERGM and multi-graph analysis, we offer evidence that structural centrality on decentralized platforms like Bluesky can reveal politically salient figures not captured by media narratives or institutional leadership roles—a finding of particular relevance for scholars of political networks and digital communication.

### 4.5 Topic Analysis

To understand the semantic contours of Congressional Democrats' discourse on Bluesky, we applied BERTopic, a transformer-based topic modeling technique, to the full corpus of posts, followed by subgroup-level modeling based on formal caucus or faction membership. Topics were extracted using a class-based TF-IDF representation and filtered by coherence and frequency thresholds.

The analysis revealed several cross-cutting themes among the top eight topics per group (Figure 4), including: cuts to Medicaid, the Israel-Gaza conflict, reproductive rights, and the Russia-Ukraine war. These topics suggest that, even on a decentralized platform like Bluesky, Congressional Democrats maintain alignment on certain high-salience policy domains. However, divergence emerges in more specialized or ideological content. For example, both the Democratic House Leadership and the Congressional Progressive Caucus featured discourse focused on Secretary of Defense Pete Hegseth, potentially reflecting concerns about recent cabinet appointments or foreign policy direction. In contrast, a unique topic surfaced among The Squad, focused on the detention of university students Mahmoud Khalil and Rümeysa Öztürk in a civil liberties context—a topic absent in other groups' top eight.

To quantify inter-group topic alignment, we computed cosine similarity between each group's top topic vectors. As visualized in Figure 5: Cosine Similarity Heatmap, the New Democrat Coalition and Congressional Progressive Caucus shared the highest topic similarity (0.4694), suggesting overlapping policy emphasis despite differing ideological branding.

Interestingly, the New Democrat Coalition also exhibited the highest similarity to the aggregate Democratic discourse (0.5485), possibly indicating a centrist positioning that aligns with mainstream party messaging. Conversely, The Squad emerged as the least similar, highlighting its distinct rhetorical and thematic priorities.

These findings reinforce earlier network-level observations: structural cohesion does not necessarily predict discourse homogeneity. While certain figures or groups may be central in terms of follower ties or mentions, their content remains meaningfully differentiated. This semantic divergence may offer insight into internal party debates and emerging factional identities post-2024.

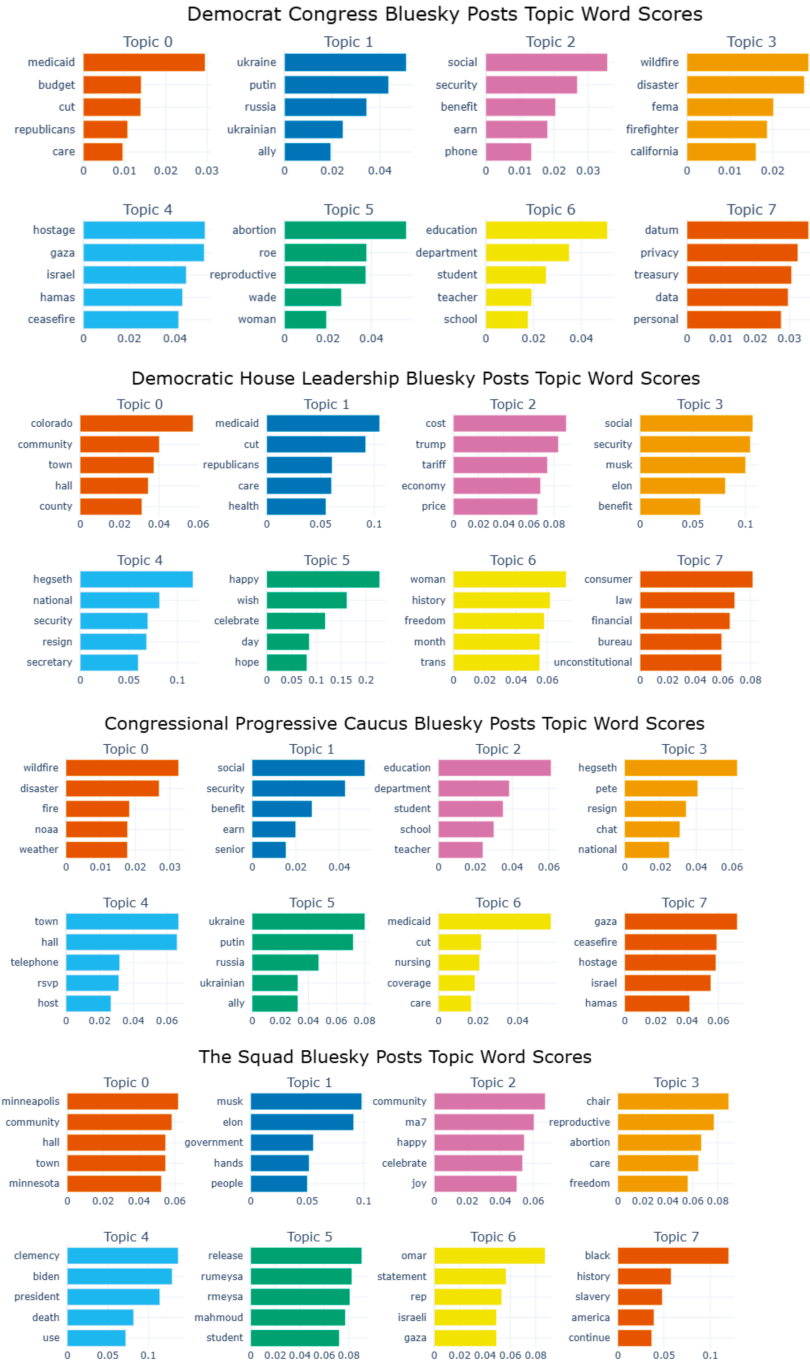

**Fig. 4.** Topic Model Visualization of Key Democrat Coalitions.

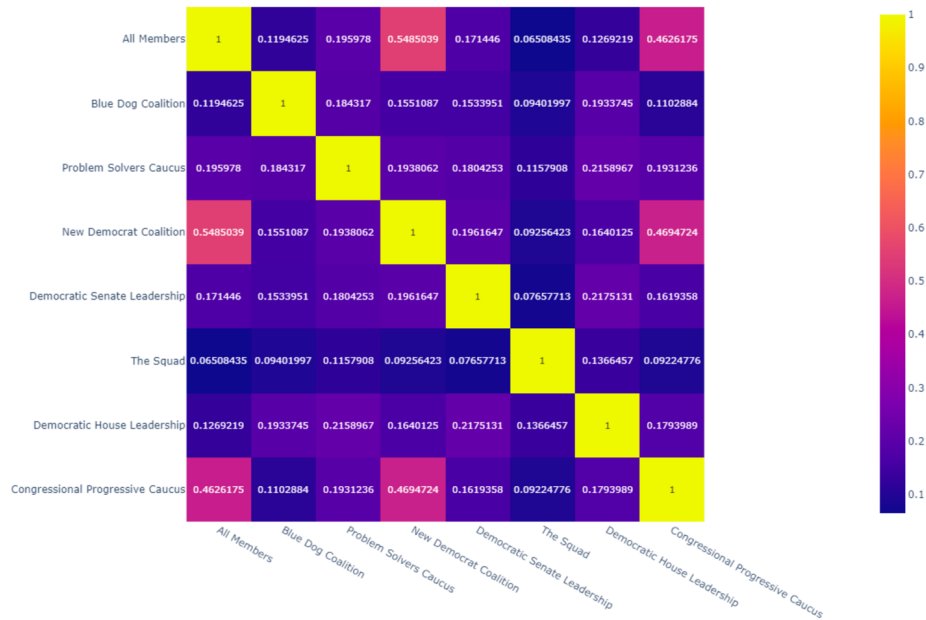

**Fig. 5.** Cosine Similarity Heatmap of Top Eight Topics for Democrat Coalitions.

# 5 Conclusion

This study provides one of the first comprehensive examinations of U.S. Congressional Democrats' activity on the decentralized social media platform Bluesky. By integrating social network analysis, exponential random graph models (ERGM), and transformer-based topic modeling, we uncover both structural and semantic patterns in elite political communication following the 2024 U.S. presidential election.

Our network analysis reveals that while prominent party leaders such as Hakeem Jeffries and Elizabeth Warren dominate follower and mention networks, several over-looked figures—such as Marcy Kaptur, Donald Beyer, and Dwight Evans—occupy structurally central positions, suggesting untapped or underrecognized influence within the party's online ecosystem. These findings underscore the value of centrality measures in identifying influential actors who may not be highly visible in media coverage or formal leadership roles.

The ERGM results provide empirical support for homophily across multiple attributes, including chamber membership, state, group affiliation, and verification status. Notably, formal leadership in the Senate appears negatively associated with tie formation and content sharing behaviors, suggesting a more hierarchical or broadcast-oriented communication style among elite figures.

Topic modeling further reveals thematic cohesion on issues such as Medicaid cuts, reproductive rights, and foreign conflicts, alongside meaningful divergence among ideological subgroups. The Squad, for example, exhibited the least topical alignment with other factions, reflecting its distinct rhetorical and advocacy priorities.

Several limitations warrant consideration. First, Bluesky remains an emerging platform with incomplete adoption among political elites; our sample omits members who have not joined or who use anonymous or unverified accounts (16.4% of congressional democrats). Second, the platform's decentralization introduces complexity in verifying authenticity and assessing cross-instance visibility. Third, while our centrality measures highlight structural influence, we do not correlate these with external popularity metrics (e.g., polling data or mainstream media citations), leaving the third research question partially open. Finally, topic modeling is inherently sensitive to preprocessing and model parameters; while BERTopic offers state-of-the-art coherence, future work could triangulate findings with alternative techniques.

Building on these findings, future research could expand in several directions. First, a comparative analysis of Democratic and Republican network structures may reveal partisan asymmetries in adoption and discourse. Note that it was difficult to verify republican accounts on the Bluesky platform. There are no House or Senate accounts on the platform with starter packs. Adding to the confusion, manual searches often yield parody accounts. Prominent republicans that we manually searched for that appear to have a valid Bluesky account like Steve Scalise and Elise Stefanik had very little activity and Stefanik appears to have deactivated her account. Perhaps comparing Republican activity on X with Democrat activity on Bluesky might be more appropriate, however, differences in access to platform data may bias findings.

Second, longitudinal data collection could track the evolution of network ties and issue salience over time, particularly as the 2026 mid-terms approach. Third, linking Bluesky behavior with off-platform outcomes (e.g., voting records, donor activity, or media attention) could provide richer insight into how digital centrality translates into real-world political influence.

Ultimately, this study contributes to our understanding of how elite political networks operate in emergent, decentralized environments. As platforms like Bluesky continue to evolve, so too must the computational tools used to analyze them—offering new avenues for research at the intersection of technology, politics, and network science.

## References

1. Ambrose, T. (2024, November 16). What is Bluesky and why are so many people suddenly leaving X for the platform? *The Guardian*. https://www.theguardian.com/technology/2024/nov/16/what-is-bluesky-and-why-are-so-many-people-suddenly-leaving-x-for-the-platform-elon-musk

2. Bonacich, P. (1987). Power and centrality: A family of measures. *American Journal of Sociology*, 92(5), 1170–1182. https://doi.org/10.1086/228631

3. Blue Dog Coalition. (n.d.). Members. *Blue Dog Coalition*. https://bluedogs-gluesenkampperez.house.gov/members

4. Bluesky. (n.d.). *About Bluesky*. Retrieved from https://bsky.social/about/faq

5.  Buzelin, A., Dutenhefner, P. R., Locatelli, M. S., Malaquias, S., Bento, P., Aquino, Y., Dayrell, L., Estanislau, V., Santana, C., Alzamora, P., Vasconcelos, M., Meira Jr., W., & Almeida, V. (2025). Analyzing Political Discourse on Discord during the 2024 U.S. Presidential Election. *arXiv preprint* arXiv:2502.03433. https://arxiv.org/abs/2502.03433

6.  Chamberlain, J. M., Spezzano, F., Kettler, J. J., & Dit, B. (2021). A Network Analysis of Twitter Interactions by Members of the U.S. Congress. *ACM Transactions on Social Computing*, 4(1), 1–22. https://doi.org/10.1145/3439827

7.  Congressional Progressive Caucus. (n.d.). Caucus members. Congressional Progressive Caucus. https://progressives.house.gov/caucus-members

8.  Freeman, L. C. (1979). Centrality in social networks: Conceptual clarification. *Social Networks*, 1(3), 215–239. https://doi.org/10.1016/0378-8733(78)90021-7

9.  Grootendorst, M. (2022). BERTopic: Neural topic modeling with class-based TF-IDF. *arXiv preprint* arXiv:2203.05794. https://arxiv.org/abs/2203.05794

10. Himelboim, I., Smith, M., Rainie, L., Shneiderman, B., & Espina, C. (2013). Classifying Twitter topic networks using social network analysis. *Social Media and Society*, 1(1). https://doi.org/10.1177/2056305113479598

11. House Democrats. (n.d.). Our members: House Democrats. House Democrats. https://www.dems.gov/who-we-are/our- members

12. Kleppmann, M., Frazee, P., Gold, J., Graber, J., Holmgren, D., Ivy, D., Johnson, J., Newbold, B., & Volpert, J. (2024). Bluesky and the AT Protocol: Usable Decentralized Social Media. *arXiv*.

13. McCulloh, I., Armstrong, H., & Johnson, A. (2013). *Social Network Analysis with Applications*. Hoboken, NJ: Wiley.

14. New Democrat Coalition. (n.d.). Members - New Democrat Coalition. New Democrat Coalition. https://newdemocratcoalition.house.gov/members

15. NPR Staff. (2024, November 19). Traffic on Bluesky, an X competitor, is up 500% since the election. *NPR*. https://www.npr.org/2024/11/19/g-s1-34898/bluesky-traffic-surge-after-election:contentReference[oaicite:22]{index=22}

16. Ocasio-Cortez, A. (n.d.). Squad. Instagram. https://www.instagram.com/p/BqGTlEPBXXD/?hl=en

17. Ozer, M., Kim, N., & Davulcu, H. (2016, August). Community detection in political twitter networks using nonnegative matrix factorization methods. In *2016 International conference on advances in social networks analysis and mining (ASONAM)* (pp. 81-88). IEEE.

18. Problem Solvers Caucus. (n.d.). Caucus Members. Problem Solvers Caucus. https://problemsolverscaucus.house.gov/

19. Sadayappan, S., McCulloh, I., & Piorkowski, J. (2018). Evaluation of Political Party Cohesion Using Exponential Random Graph Modeling. In *Proceedings of the 2018 IEEE/ACM International Conference on Advances in Social Networks Analysis and Mining (ASONAM)* (pp. 1–8). IEEE.

20. Senate Democrats. (n.d.). Senate Democrats Starter Pack. Bluesky Social. https://bsky.app/starter- pack/democrats.senate.gov/3ljiq7krz5223

21. UnitedStatesSenate.(n.d.).Leadership&Officers.UnitedStatesSenate.https://www.senate.gov/senators/leadership.htm

22. United States Senate. (n.d.). Senators. United States Senate. https://www.senate.gov/senators/index.htm

23. U.S. House of Representatives. (n.d.). Leadership. United States House of Representatives. https://www.house.gov/leadership

24. U.S. House of Representatives. (n.d.). Representatives. United States House of Representatives. https://www.house.gov/representatives